# OPTICS Based Coverage in Wireless Sensor Network

Chandranath Adak

Dept. of CSE, University of Kalyani, West Bengal-741235, India
mr.c.adak@ieee.org

**Abstract.** This paper deals with the coverage problem of wireless sensor network. We use the density based clustering technique - OPTICS to cover a target region with less number of sensor nodes. OPTICS works well to identify the outliers, core points and it obtains the denser regions. We define a level of acceptance to find next appropriate sensor in the region. We eliminate overlapped area and obtain a decision tree to minimally cover up the target region.

## 1   Introduction

*Wireless Sensor Networks* (WSN) [1] are used in wide range of prospective applications such as surveillance, object tracking, natural security, information broadcasting, biological detection, environmental monitoring, weather reporting, underwater and remote sensing etc. over the last two decades. WSN is composed with tiny, low-power, cheap radio devices (sensors) to sense and communicate with server. The aim of coverage problem [2-4] is to use minimum number of sensors and to cover the entire target region. Sometimes full coverage is not required, partial coverage [5] fulfils the necessity.

Meguerdichian et al. [6] introduced coverage problem in ad-hoc WSN. Jourdan and Weck [7] proposed multi-objective genetic algorithm (MOGA) based layout optimization. Hongli Xu et al. [8] used localized algorithm for coverage in WSN. Liu and Liang [9] worked with the approximate coverage. Zou and Chakrabarty [10] proposed a distributed coverage and connectivity centric technique to select active nodes. Wang, Wu and Guo [11] presented k-coverage in 3D WSN. Almuzaini and Gulliver [12] proposed range based localization by DBSCAN in wireless network. Yoon and Kim [13] deployed maximum coverage in WSN using genetic algorithm.

   Here we use the OPTICS [14] algorithm to find initial arbitrary shaped clusters. As it works fine to recognize outliers and core points, system can easily visualize the dense regions to work at a priority. Now we define a acceptance level (depending on a sensor's battery life, direct neighbors and distance measurement) to find the coverage in all clusters concurrently. We discard the overlapped coverage area of the sensors. Working with all clusters and making a decision tree [16], we find minimum number of sensors to cover the clusters, as well as the target region.

## 1.1 OPTICS: Ordering Points To Identify the Clustering Structure

*Clustering* is a technique to divide a data set into subsets (clusters) by their similarities (or dissimilarities). The pattern within/outside a cluster is homogeneous/heterogeneous.

*Density based clustering* separates the denser region (number of data objects within a neighborhood with respect to a threshold) from the low-density region. It can find non-globular (arbitrary) shaped clusters and works fine to filter out the outliers (significantly deviated data objects).

*OPTICS* [14] is a density-based clustering method. Instead of producing clustered data set, it produces a linear object list to represent the density-based clustering structure. Data objects in denser region are listed closer. This linear structure is easily analyzed to extract the information about cluster centers, outliers and arbitrary-shaped clusters. Due to these advantages, we use OPTICS to deal with coverage problem in WSN. OPTICS does not require user-specific density threshold, it needs *core distance* and *reachability distance*. The complexity is $O(nlogn)$ for spatial indexing and $O(n^2)$ otherwise, where $n$ is the number of objects.

**ε:** maximum radius of a neighborhood,
**MinPts:** minimum number of neighborhood objects of a *core point*,
**core point:** ε-neighborhood of an object contains at least *MinPts* of objects,
**core distance:** minimum distance threshold to make object $p$ a *core point*,
**reachability distance:** $p$ and $q$ be two objects. $p$ is a core point and $q$ is the neighborhood of $p$. The reachability-distance from $p$ to $q$ is *max{core-distance(p), dist(p,q)}*.

---

**Algorithm 1** The OPTICS Algorithm

1: *OrderSeed*: output ordering object list, sorted by *reachability_distance* from closest *core_point*.
2: **while** (!empty(*inputDataSet*) || !empty(*OrderSeed*)) **do**
3:     choose arbitrary object $p$ from *inputDataSet*;
4:     find *eps_neighborhood* of $p$ and determine *core_distance*;
5:     *reachability_distance*=undefined;
6:     **if** (p==*core_point*) **then**
7:         find each object $q$ in *eps_neighborhood* of $p$;
8:         update *reachability_distance* of $q$ from $p$;
9:         **if** (!processed($q$)) **then**
10:           insert $q$ into *OrderSeed*;
11:         **end if**
12:     **else**
13:         **if** (!empty(*OrderSeed*)) **then**
14:           move to next object in *OrderSeed*;
15:         **else**
16:           move to next object in *inputDataSet*;
17:         **end if**
18:     **end if**
19: **end while**

The data points are plotted through *clustering order* (horizontal axis) with its *reachability distances* (vertical axis). The *Gaussian bumps* in the plot denote number of clusters in dataset.

## 2  Proposed Method

The proposed method consists of following steps:

1. Suppose there are *n* numbers of sensors in a target region. These sensors are treated as object points; we cluster these *n* object points using OPTICS algorithm and get *k* clusters. [ $0 < k \leq n$ ].

2. Choose initial sensor node $S_i$ from each cluster.

3. $S_i$ broadcasts *REQ* (request signal) to its neighborhoods $S_{j[1,...,m<n]}$, where $dist(S_i , S_j ) \leq 2r \leq 2\varepsilon$ . [ *r*: coverage range of a sensor, assume radius *r* is same for all sensors]. We use *Euclidian distance function* to measure $dist(S_i , S_j )$.

4. $S_i$ receives *ACK* (acknowledgment signal) from $S_j$ whose $L_j$ is maximum.
   $L_j = (0.4*B_j + 0.3*N_j)/(0.2*D_{ij})$, where

   $L_j$ : level of acceptance,
   $B_j$ : battery-life of $j^{th}$ sensor node,
   $N_j$ : number of direct neighbor of $j^{th}$ node (*Appendix*: fig.1),
   $D_{ij}$: $dist(S_i , S_j )$, i.e. Euclidian distance between sensor node $S_i$ and $S_j$.

   This $L_j$ gives the measurement to choose the sensor that has higher battery-life, more direct neighbors and which is closer to $S_i$.

5. For overlapping neighbor, we discard the overlapping region of coverage range and work with the non-overlapped perimeter, coverage area [4] (*Appendix*: fig.2).
   Non-overlapped perimeter = $2r(\pi-\alpha)$.

6. Steps *2-5* are repeated to make a *decision tree* within a cluster and to hierarchically cover a clustered region with minimum number of sensors.

7. All the cluster with minimum number of sensor nodes are combined together to cover the target zone.

8. After After a specific time period these active nodes [10] (working sensors right now) go to *sleep mode* and inactive nodes follow steps 2-7 to cover the region.
   Steps *2-8* run periodically.

## 3 Experimental Result

We have implemented our proposed method in MATLAB-R2012a [15]. The number of active nodes does not remain constant with changes in deployed nodes (D); and the number of active nodes ($n_1$, $n_2$, $n_3$) varies due to arbitrary change in cluster for a particular number of deployed nodes.

The deployed area is fixed, i.e. *50X50 $m^2$*. We take *100-500* number of deployed nodes (sensors with *r=5m*) for our test case analysis (table.I). The average number of active nodes (*N*) is calculated by taking arithmetic mean of $n_1$, $n_2$ and $n_3$.

**Table I.** *Result Analysis*: #Deployed Node vs. # Active Node

| # Deployed Node (*D*) | # Active Node | | | | R = [*(N/D)*100*] % |
|---|---|---|---|---|---|
| | $n_1$ | $n_2$ | $n_3$ | N= [($n_1$+$n_2$+$n_3$)/3] | |
| 100 | 27 | 32 | 40 | 33 | 33 |
| 150 | 49 | 51 | 62 | 54 | 36 |
| 200 | 53 | 57 | 67 | 59 | 30 |
| 250 | 58 | 63 | 86 | 69 | 28 |
| 300 | 77 | 82 | 87 | 82 | 28 |
| 350 | 91 | 95 | 104 | 97 | 28 |
| 400 | 109 | 127 | 130 | 122 | 31 |
| 450 | 121 | 128 | 152 | 134 | 30 |
| 500 | 140 | 152 | 163 | 152 | 31 |

$R_{avg}$=⌈*average(R)*⌉ % = ⌈30.55⌉ % = ***31 %***.

We measure *coverage ratio (CR)* [17] (ratio of total coverage area by all sensors and size of the target area) to check the performance of our proposed method.

$$CR = \frac{\pi * (5m)^2 * 31}{50 * 50 m^2} * 100\% = \mathbf{97.389\%}$$

On the basis of our test case analysis, our proposed method covers 97.389% of the target region with 31% of deployed nodes (sensors).

We compare our method with some of the existing methods in table.II.

**Table II.** Comparison of Results

| Sl. No. | Methods | Coverage Ratio (≈ %) |
|---|---|---|
| **1.** | RAND (random) [15] | 62 |
| **2.** | Fuzzy C-Means (FCM) | 74 |
| **3.** | Z. Tong et al. [18] | 87 |
| **4.** | SA-MODE [19] | 92 |
| **5.** | Proposed | 97 |

## 4  Conclusion

We use OPTICS to deal with coverage problem in WSN. The experimental result shows well performance of the proposed method, i.e. 97.389% of the target region can be covered up with 31% of deployed nodes. Due to use of clustering technique, the object sets are chosen arbitrarily. So, the possibility of getting new active sensors in periodic time is high. Though there is a little bit chance to neglect an important sensor, and choose that as an outlier due to randomization of our proposed method, it works fine due to well performance of OPTICS in elimination of outliers.

## Appendix:

$N_j$ : number of direct neighbor of $j^{th}$ sensor node ($S_j$)

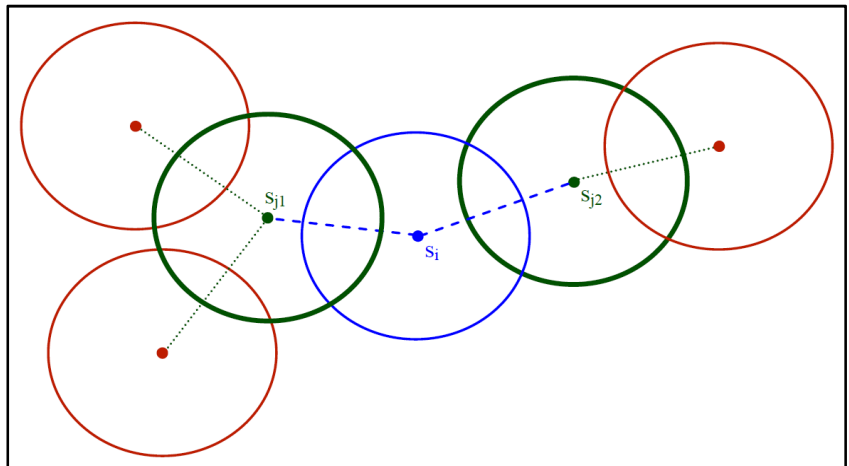

Fig. 1. Illustration of direct neighbor of a sensor

From fig.1. , $S_{j1}$ has two direct neighbors and $S_{j1}$ has one direct neighbor.
So, $N_{j1} = 2$ and $N_{j2} = 1$ .

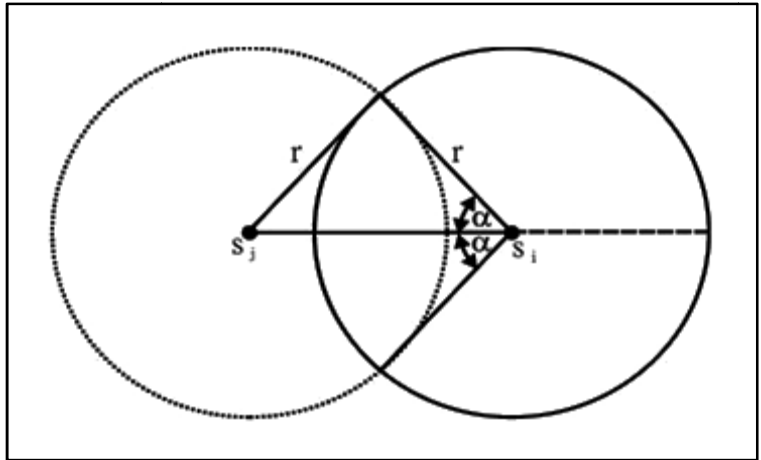

Fig. 2. Overlapping of sensor regions

From fig.2., Total perimeter of sensor region = $2\pi r$
Overlapped perimeter of $S_i$ = $r.2\alpha = 2\alpha r$
Non-overlapped perimeter of $S_i$ = $2\pi r - 2\alpha r = 2r(\pi - \alpha)$